\documentclass[12pt]{article}
\usepackage[dvips]{graphicx}
\usepackage{pdproc}

  \makeatletter 
  \def\@cite#1{[#1]} 
  \makeatother    
\begin{document}

\renewcommand{\thefootnote}{\alph{footnote}}

\title{
Minimal Flavor Violation at Large Tan$\beta$ 
}

\author{CHRISTOPHER KOLDA}

\address{ 
Department of Physics, University of Notre Dame\\
Notre Dame, Indiana 46556, USA
}

\abstract{I review briefly the notion of minimal flavor violation and
  its application to supersymmetry, with a special emphasis on a class
  of operators, so-called Higgs penguins, which can generate new and
  interesting flavor signals at large $\tan\beta$. \\~~\\
  {\sl (Talk presented at
  the 12th International Conference on Supersymmetry
  and Unification of Fundamental Interactions (SUSY '04), 
  June 2004, Tsukuba, Japan. To be published in the proceedings.)}}

\normalsize\baselineskip=15pt

\section{Care and Feeding of the Higgs Penguin}

If it were not for the gauge hierarchy problem, there would be no
reason to expect the Standard Model (SM) of particle physics to be
modified at scales near those being explored today. Extensions of the
SM at the weak scale face such high hurdles that one could easily
argue that new physics simply will not be found there. 
These hurdles include precision
measurements of the electroweak oblique parameters and the 
non-observation of flavor-changing neutral currents (FCNCs), CP
violation, or baryon number violation at levels that one would expect from
most models of new, weak-scale physics. 

Nonetheless the gauge hierarchy problem needs a solution, and it needs
a solution at the weak scale. This tension is the central problem in
theoretical particle physics right now. The best solution to the
hierarchy problem, supersymmetry (SUSY), is only partially successful
in overcoming the constraints listed above. In particular, it has
great difficulty with the FCNC and CP violation constraints. One would
not expect 
``generic'' new physics which couples non-universally to flavor (such
as SUSY) below scales of 10's to even 1000's of TeV. Yet SUSY must be
below 1~TeV in order to solve the hierachy problem.

In generic new models of physics, there is one very attractive approach to
preventing large new flavor signals: we can require that the model be
{\it minimally flavor violating} (MFV). 
Such a condition has been defined in many
ways over the years, but I have in mind here the definition proposed
by Ref.~\cite{mfv} in which one imposes a global $SU(3)^5$ flavor
symmetry on the particle spectrum and interactions, under the added
assumption that only the Yukawa matrices violate the symmetry. In such a
picture a spurion analysis reveals exactly what kinds of FCNC signal
can and cannot show up.

Usually one can make a statement in MFV models that no new operators
arise which are not already in the SM and that new contributions to 
SM operators are suppressed by the same CKM factors that suppress
the SM contributions. Thus new MFV physics changes SM
predictions for FCNC processes by $O(1)$ at best. A very special case
is CP-violating asymmetries where one can go farther and show that MFV
models cannot change these at all! The minimal SUSY Standard Model
(MSSM) is not necessarily an MFV
theory, but certain classes of MSSM ``models'' fit the mold. These include
models with unified scalar masses (mSUGRA models), anomaly-mediated
and gauge-mediated models. For example, in the mSUGRA
models flavor violation only enters through the
renormalization group equations (RGEs). For example, the left-handed
scalar mass RGEs have a general form:
\begin{equation}
\frac{d}{dt}\left(m^2_{\tilde Q}\right)_{ij} \propto a{\bf 1}_{ij}
+b\left(Y_U^\dagger Y_U\right)_{ij}
\end{equation}
where the up-quark Yukawa matrix, $Y_U$, is non-diagonal.

However the claim that no new operators arise, and that all SM-like
operators have a SM-like flavor suppression, does not hold in the MSSM.
There is a hidden
assumption underlying the original claim: that the low-energy effective theory
is the {\it minimal}\/ Standard Model; that is, it contains only a
single Higgs field. If the low energy theory is a two Higgs doublet
model, then new operators arise which can generate large, new
FCNCs. Such is the low energy limit of the MSSM
when the second Higgs doublet is not too heavy ({\it i.e.}, $m_A\,\,\,
\slash\!\!\!\!\!\!\!\gg m_Z$).

How does this show itself? It is convenient to work in a basis in which
\begin{equation}
Y_D=\mbox{diag}(y_d,y_s,y_b),\quad Y_U=V^\dagger \times \mbox{diag}
(y_u,y_c,y_t)
\end{equation}
where $V$ is the usual CKM matrix. The leading flavor changing
operators must all scale as $(Y_U^\dagger Y_U)_{ij}\propto y_t^2
V_{3i}^* V_{3j}$. There could also be operators involving powers of
$Y_D$, but in a single Higgs model these are always suppressed by
powers of $m_b/m_t$.

But because the MSSM is a two Higgs model, the suppression is instead
$(m_b/m_t)\tan\beta$, 
which can be $O(1)$ if $\tan\beta$ is large. For example, a
one-Higgs model would never generate a large $\bar b_R s_L$ operator
because it would always be suppressed by $m_b$. But in the MSSM, it is
instead suppressed by $m_b\tan\beta$, which need not be a suppression
at all.

It is by now well known that in the MSSM such operators can arise and
be important. New flavor-conserving $\bar b_r b_L$ 
operators were discussed most explicitly by Hall {\it et
  al.}~\cite{hrs}, though there existed a long
prehistory in which a number of authors took note of their
existence~\cite{pre}. The flavor-violating operators were discovered
and studied much later. The discussion here is based on the paper of
Babu and Kolda~\cite{bk}, though other early related discussions can
be found in Refs.~\cite{begin}. (A recent review, with a more complete
list of references, can be found in Ref.~\cite{dedes}.)

There are two approaches commonly used to calculate the new FCNC
operators: the effective Lagrangian approach, or a Feynman
diagrammatic approach. The former has the advantage of resumming the
leading contributions from several orders of diagrams and is process
independent. The latter, though more unwieldy, demonstrates the new
contributions explicitly in diagrams which are reminiscent of penguin
diagrams, except with a Higgs boson replacing the vector boson. Thus
the name ``Higgs penguins'' has arisen to describe this new physics.
Nonetheless we will follow the simpler effective Lagrangian approach
pioneered in Ref.~\cite{bk}.

In the effective Lagrangian approach one integrates
out the SUSY partners (but not the second Higgs doublet) at the scale
$M_{\rm SUSY}$. In doing so, an effective $\bar d_{Ri} d_{Lj} H_u^c$
interaction is generated. For $i\neq j$, there are two leading
diagrams that must be included: a 1-loop Higgsino diagram and a 2-loop
gluino diagram (see Fig.~1). Both generate the new operator $\bar Q_L
(Y_U Y_U^\dagger) Y_D D_R H_U^c$. The Higgsino diagram generates a
coefficient of order $1/(16\pi^2)$, while the gluino diagram generates
a coefficient of order $\alpha_s/(9\pi^3)\log(M_X/M_{\rm SUSY})$. If
the log is large, then this 2-loop contribution can be as important as
the 1-loop piece.

\begin{figure}[t]
\begin{center}
\includegraphics*[width=4cm]{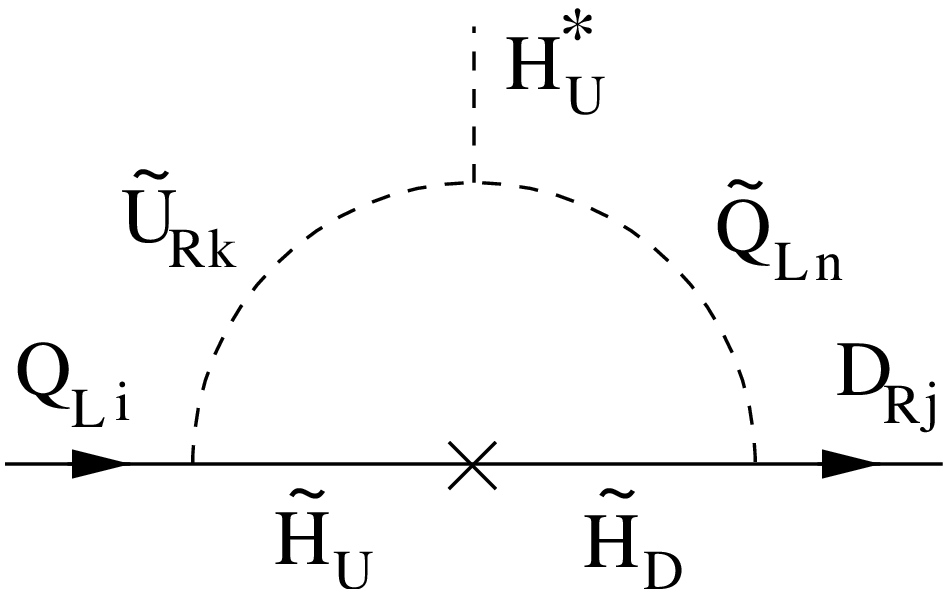}
~~~~~~~
\includegraphics*[width=4cm]{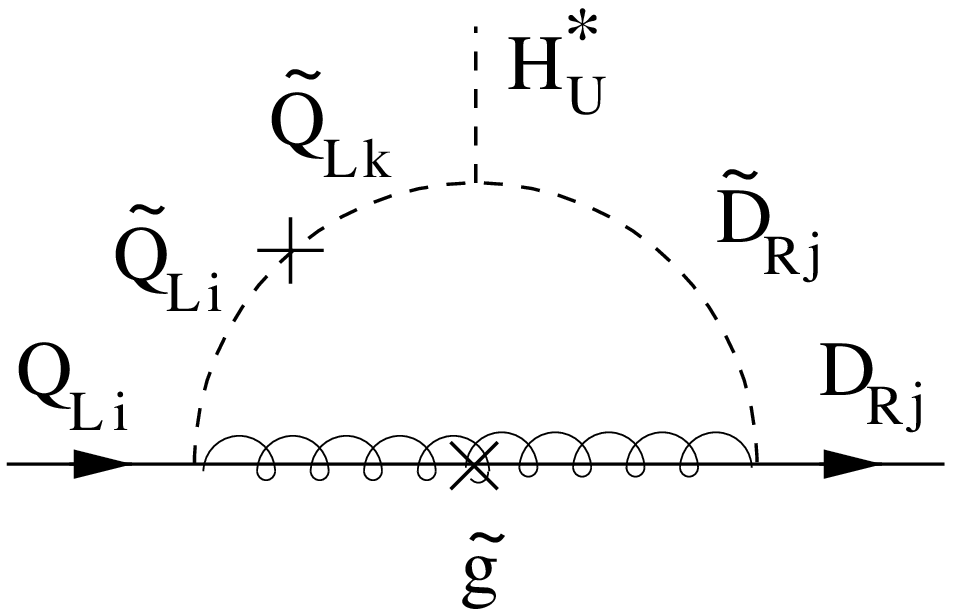}
\caption{%
The two leading diagrams in the calculation of the effective $\bar b_R
s_L H_u^c$ interaction. The cross in the second diagram represents a
one-loop flavor mixing contribution from the RGEs.
}
\label{fig1}
\end{center}
\end{figure}

When the $H_u$ field gets a vev, these operators generate new
contributions to the $d$-quark mass matrix. For example, keeping the
$s$- and $b$-quarks only, one has a mass matrix
\begin{equation}
-{\cal L}=\left(\begin{array}{cc} \bar s_R & \bar b_R \end{array}
\right) \left(\begin{array}{cc} y_s v\cos\beta & 0 \\ \epsilon
y_b y_t^2 v\sin\beta & y_b v\cos\beta \end{array}\right)
\left(\begin{array}{c} s_L^0 \\ b_L^0 \end{array}\right)
\end{equation}
where $\epsilon$ is the sum of the two contributions discussed
previously. Diagonalizing this matrix requires the $s_L^0$ and $b_L^0$
to mix at an angle
\begin{equation}
\sin\theta \simeq \epsilon y_t^2 \tan\beta
\end{equation}
which can be $O(1)$ is $\tan\beta$ is large. Thus we have a brand new
flavor-changing parameter, $\sin\theta$, which does not correspond to
any parameter of the SM.

Where is this new physics important? The biggest effect comes in
purely leptonic $B_s$-meson decays through the operator
$(\bar b_R s_L)(\bar\ell_R\ell_L)$. Because of the chiral structure on
both sides, this operator must scale as $y_by_\ell$, which implies a
$\tan^2\beta$ dependence to the operator. But the angle $\theta$ has
its own $\tan\beta$ dependence, so the entire operator scales as
$\tan^3\beta$. Of course, the operator is dimension six and so must
fall as $1/\Lambda^2$. Here we can interpret $\Lambda=m_A$ since in
the limit $m_A\to\infty$, our model contains only a single Higgs and
is therefore immune to new operators. For the remainder of this talk I
will review what is known about the leading channels for discovering
this SUSY ``violation'' of minimal flavor violation, and what we may
learn from them.

\section{Experimental Signatures}

${\bf B_{s,d}\to\mu\mu}$: This is the gold-plated mode. The SM rates
are GIM- and helicity-suppressed (B$(B_s\to\mu\mu)\simeq 4\times 10^{-9}$),
so the SUSY rate can be several
orders of magnitude larger than the SM. This mode is easily studied at
the Tevatron where plentiful $B_s$ mesons are produced and their
decays to muons are easy to tag. In fact, this may provide our best hope for
finding SUSY at the Tevatron, as the Tevatron is sensitive to rates
down to about $10^{-7}$ over the next few years.
On the other hand $B$-factories, which make $B_d$
but not $B_s$, suffer a $(V_{td}/V_{ts})^2$ suppression in their rates,
which is enough to put them outside the interesting range at present.

~

${\bf B_{s,d}\to\tau\tau}$: While this is the ideal mode
theoretically, beating $B\to\mu\mu$ by $(m_\tau/m_\mu)^2$, 
it is not a very good channel experimentally. This could be studied at
a $B$-factory with a large sample of fully reconstructed $B_d$'s, but
that would require SuperB factory luminosities.

~

${\bf B_d\to X_s\mu\mu}$: In this 3-body decay, the SM contribution is
no longer helicity suppressed, eliminating one of the big advantages of the
$B_s\to\mu\mu$ mode. However SUSY can still generate $O(1)$ corrections
to the branching fraction. In order to cancel the large uncertainties
in the SM rate calculation, Hiller and Kr\"uger~\cite{hiller} have 
suggested a search
for deviations in $\mbox{B}(B_d\to K\mu\mu)/\mbox{B}(B_d\to K
ee)$. With the appropriate kinematic cuts, the SM prediction for this
ratio is known to better than 0.01\%. Though current experimental data 
does not constrain the Higgs penguin as strongly as
$B_s\to\mu\mu$, it is approaching that level of utility. Plus it
provides information from the $B_d$ sector which would be a nice
consistency check on the Higgs penguin picture.

~

{\bf ${\bf B^0-\overline{B}^0}$ mixing}: In MFV models, even those with two
Higgs doublets, this can be shown to be suppressed beyond naive power
counting~\cite{mfv}. 
However contributions do arise at $O(m_s/m_b)$~\cite{buras} 
and at two loops~\cite{bk}. Buras {\it et al.}~\cite{buras} have found that,
given CDF bounds in $B_s\to\mu\mu$, the Higgs penguin contributions to
$B_s$--$\bar B_s$ mixing could suppress $\Delta M_s$ by as much as 50\%
from its SM prediction. They also found the suppression to be highly
correlated with the rate for $B_s\to\mu\mu$, suggesting another nice
consistency check on the Higgs penguin scenario.

~

There is no single definition of minimal flavor violation in the
leptonic sector. Applying the same logic from the quark sector would
lead to no flavor violation at all. However the evidence of neutrino
mass and mixing leads many theorists to posit the existence of
right-handed neutrinos at a very large mass scale. With these
neutrinos comes a new Yukawa interaction and its accompanying coupling
matrix. Though the scale of the new right-handed Yukawa interactions
would be far above $m_W$, many authors have noted that in SUSY the
scalar mass spectrum will still bear an imprint from those
interactions thanks to the renormalization group.

In a typical case, one supposes a highly non-diagonal Yukawa
matrix, with several $O(1)$ entries,
for the right-handed neutrinos at scales around
$10^{14}\,$GeV. In such a case, one finds low energy evidence for the
interaction through a variety of channels, particularly
$\tau\to\mu\gamma$ and $\mu\to e\gamma$, both mediated by gauginos. 
However there are also related Higgs-mediated processes which will violate
lepton number, including:

~

${\bf \tau\to 3\mu,\mu ee}$: The leptons that appear in the final
state can tell us which entries in the neutrino Yukawa matrix are
large and which aren't~\cite{bk2}.

~

${\bf B_{s,d}\to \ell\ell'}$: A Higgs-mediated FCNC $+$ 
Higgs-mediated lepton flavor violaton $=$ the ultimate Higgs penguin
process!~\cite{ellis}

~

${\bf H^0,A^0\to\ell\ell'}$: Once the heavy Higgs bosons are found,
this could be a large effect since it does not suffer from Higgs
decoupling effects~\cite{brignole}. 

\section{A Useful Probe of Fundamental Physics}

A number of authors have commented on the utility of the Higgs penguin
for extracting information on more fundamental questions in SUSY and
beyond. In fact much can be learned about SUSY 
from these rare, Higgs-induced processes even without seeing a single
superpartner! For example:
\begin{itemize}
\item Gauge-mediated SUSY breaking models produce little observable
  signal of a Higgs penguin. Observation of $B_s\to\mu\mu$ at the
  Tevatron would certainly rule these models out~\cite{baek}.
\item Supergravity-type models can produce large signals at the
  Tevatron, strengthening their position if $B_s\to\mu\mu$ is seen.
\item We can place model-independent bounds on $\tan\beta$ and
  $m_A$~\cite{kkl,dedes2} if Higgs penguin effects are observed.
\item In the leptonic sector, any channel in which lepton flavor
  violation is observed corresponds directly to an $O(1)$ element in
  the neutrino Yukawa matrix -- a matrix which may decouple at
  $10^{14}\,$GeV! 
\end{itemize}
The flavor sector has often been a source of consternation among
theorists working in SUSY, but the Higgs penguin is one example where
flavor physics 
may provide unique opportunities for probing fundamental physics at
a deeper level. If $\tan\beta$ is large, there may be many
opportunities to study the Higgs penguin in the next few years and
many ways in which it will further our understanding of the electroweak
scale and beyond.

\bibliographystyle{plain}

\end{document}